\documentclass[fleqn,10pt]{wlscirep1}

\usepackage{graphicx}
\usepackage{bm}

\title{Easily doped $p$-type, low hole effective mass, transparent oxides}

\author[1,*]{Nasrin Sarmadian}
\author[1]{Rolando Saniz}
\author[1]{Bart Partoens}
\author[2]{Dirk Lamoen}
\affil[1]{CMT, Departement Fysica, Universiteit Antwerpen, Groenenborgerlaan 171, B-2020 Antwerpen, Belgium.}
\affil[2]{EMAT, Departement Fysica, Universiteit Antwerpen, Groenenborgerlaan 171, B-2020 Antwerpen, Belgium.}
\affil[*]{nasrin.sarmadian@uantwerpen.be}

\begin{document}
\maketitle

\thispagestyle{empty}

\section*{Abstract}

Fulfillment of the promise of transparent electronics has been hindered until now largely by the lack of semiconductors that can be doped $p$-type in a stable way, and that at the same time present high hole mobility and are highly transparent in the visible spectrum. Here, a high-throughput study based on first-principles methods reveals four oxides, namely X$_2$SeO$_2$, with X$=$La, Pr, Nd, and Gd, which
are unique in that they exhibit excellent characteristics for transparent electronic
device applications -- i.e., a direct band gap larger than 3.1 eV, an average hole effective mass below the electron rest mass, and good $p$-type dopability. Furthermore, for La$_2$SeO$_2$ it is explicitly shown that Na impurities substituting La are shallow acceptors in moderate to strong anion-rich growth conditions, with low formation energy, and that they will not be compensated by anion vacancies $V_{\rm O}$ or $V_{\rm Se}$.


\section*{Introduction}

Although $p$-type transparent conducting oxides (TCOs) have
already made their way into optoelectronic devices such as
light emitting diodes (LEDs), transparent thin film transistors,
and solar cells,
\cite{app_p1,app_p2,app_p3} current $p$-type TCOs are unable to match the performance of $n$-type TCOs in terms of charge carrier mobility and/or optical
transparency \cite{app_p1,hosono07,nagarajan01,delafossite}.
This has significantly
interfered with device efficiency or with the development
of advanced applications based on active transparent
electronic components, such as bipolar transistors or diodes.
Thus, it is of great interest to find new $p$-type TCO compounds, with
properties that would lead to a technological breakthrough
and herald the real era of transparent
electronics \cite{wager08}.
Still, in spite of considerable
research efforts for more than a decade, this goal has not yet
been reached \cite{app_p1,hosono07,facchetti10}.
The data on established $p$-type TCOs
compiled in Table~2.1 of Ref.~\citenum{wager08} (page 10)
are quite illustrative.
Compounds exhibiting
a good transparency (above 70\%), tend to show very
low conductivities, typically below 1 S cm$^{-1}$, while those
with a conductivity above 10 S cm$^{-1}$ have a poor optical
transparency (below 30\%). In comparison, Sn-doped In$_2$O$_3$,
the most widely used $n$-type TCO, has an optical transparency
above 80\% and a conductivity of the order of
10$^4$ S cm$^{-1}$ \cite{edwards04}.
The two intrinsic properties of a host
TCO material that are critical in this regard are its band gap
and the charge carrier effective mass. The former should be
wide enough that visible light (frequencies ranging roughly
from 1.59 to 3.18 eV
\cite{hecht02}) is not absorbed
and the latter should be conducive to a good charge carrier mobility. In the case of In$_2$O$_3$,
the band gap is of 3.75 eV
and the electron effective mass of
$\sim0.35$ $m_e$ (with mobilities ranging from 20 to
50 cm$^2$V$^{-1}$s$^{-1}$, depending on carrier density)
\cite{edwards04}.
However, it is important to realise that a wide band gap and a
low hole effective mass do not readily
promise a good $p$-type TCO.
A clear example of this is given by ZnO (a very good
$n$-type TCO when doped, e.g., with
aluminum \cite{janotti09}).
Indeed, the hole effective mass
values for ZnO range from 0.31 $m_e$ to
0.59 $m_e$, depending on direction and band
\cite{morkoc09}.
For this reason, and because of its good transparency, there have been continuous attempts to obtain $p$-type doped ZnO. These
efforts have resulted in $p$-type conductivity in doped ZnO
being announced several times in the past, even with mobilities
rivaling those of $n$-type TCOs \cite{janotti09}.
Unfortunately,
it was invariably found
later on that the $p$-type conductivity was unstable,
i.e., the acceptors were eventually compensated, typically
by donor native defects \cite{janotti09}.

The above underlines the notorious difficulty in finding a
transparent oxide that has a low hole effective mass and that
at the same time can be doped $p$-type in a stable way. The
reasons for this are fairly understood and have been discussed
by previous authors \cite{hosono07,zunger03}.
Briefly, the states
around the valence
band maximum (VBM) in oxides are typically of oxygen $2p$
character, which has two implications. First, these states
tend to be localised, so the dispersion
around the VBM is low and, consequently, the hole effective
mas is large. Second, the VBM tends to
be deep below the vacuum level, i.e., the
ionisation potential is large. Thus, finding shallow
acceptors is difficult and/or these will tend to be
compensated. To solve this conundrum
one may consider alloying with an element with $3d$ orbitals close
to the oxygen $2p$ orbitals, so that hybridisation lifts
the VBM, making $p$-type doping feasible and lowering the
effective hole band mass. This was the original idea behind
the first $p$-type TCO to be produced, CuAlO$_2$ \cite{hosono07}.
Another idea, again put forward by Hosono and co-workers
\cite{hosono07}, is to replace oxygen with a chalcogen
(S, Se, or Te), which have more delocalised $p$ orbitals.
This led to the discovery of LaCuOS and LaCuOSe as $p$-type
oxides \cite{hosono07}. However, in the former mobility is
poor, while in the latter transparency is insufficient
because of its smaller gap \cite{ueda00,hiramatsu03}.
Recently, thanks to their high-throughput
work, Hautier and co-workers \cite{7b} found that
the presence of pnictogens, such as P and As, can result
in a low hole mass as well.
They also found that this can occur if
the oxygen $2p$ orbitals can hybridise with $s$ orbitals, in addition to closed shell $d$ orbitals [i.e., $(n-1)d^{10}ns^2$
orbitals]. This is the case,
for instance, of A$_2$Sn$_2$O$_3$ (A=K, Na), K$_2$Pb$_2$O$_3$, and PbTiO$_3$. However, a theoretical study based
on the $GW$ approximation \cite{GW1,GW2} (calculation of the eigenvalues based on many body perturbation theory) indicates that, with a band gap
$\leq 2.6$ eV, the first three
present an insufficient transparency \cite{7b}.
On the other hand, following the same approximation, the
band gap of
PbTiO$_3$ was found to be 3.7 eV, but its stable $p$-type
dopability is uncertain \cite{7b}.

In light of all the
research efforts mentioned, the question is how to proceed to
try to
identify, as efficiently as possible, materials with a wide
enough band gap, a low hole effective mass, and which
can be easily doped $p$-type, i.e., in which $p$-type
conductivity will be stable. Indeed,
given the very large number of existing oxides that can be studied, or possible chemical modifications that can be made,
a systematic experimental study is not possible. High-throughput {\it ab initio} computations, on the other hand, can be used to screen large classes of materials, searching for those that exhibit a predetermined basic set of properties, qualifying them as potential candidates for a specific
application \cite{5d,ch4_4}. This approach has already been used in the search for novel organic $p$-type semiconductors
(not transparent) \cite{5} and candidate TCO materials \cite{7b,bixbyite}, as well as new thermoelectric \cite{ch4_6},
piezoelectric \cite{t13} and scintillator materials
\cite{aflowlib_3}.

In this work we use a high-throughput search engine
to screen all the binary, ternary, and quaternary oxides reported in the AFLOWLIB computational database
\cite{aflowlib_2,aflowlib_r2} (12211 oxides in all),
and make a first selection of compounds
with a wide band gap and low hole effective
mass. Systematic
higher-accuracy first-principles electronic structure
calculations
allow us subsequently to obtain
a final list of four
oxides that we predict to be easily doped
$p$-type, while being
transparent nominally in the entire visible range and
having an effective hole mass lower than 1 $m_e$,
namely La$_2$SeO$_2$, Pr$_2$SeO$_2$,
Nd$_2$SeO$_2$, Gd$_2$SeO$_2$.
For demonstration, we consider La$_2$SeO$_2$ and
show that Na impurities substituting
La (Na$_{\rm La}$) in this material are shallow acceptors
in moderate to strong anion-rich
conditions, with a low formation energy,
and that they will not be compensated by anion
vacancies $V_{\rm O}$ or $V_{\rm Se}$.

\section*{Results}

\subsection*{Database screening}

The AFLOWLIB is an extensive repository of computational data
on materials, including structural and
electronic properties. We wrote a Python-based search engine to screen the hole effective masses and band gaps
of all the binary (1885), ternary (6416),
and quaternary (3910) oxides in the AFLOWLIB database, all
of them completely identified compounds from
the Inorganic Crystal Structure
Database (ICSD) \cite{ICSD}. To increase the chances
that our final compounds will present a comparably
good hole mobility, we select those oxides with an
average effective hole mass $\leq 1$ $m_e$, which is
considerably lower
than in most of the current $p$-type TCOS
\cite{delafossite,hiramatsu07}. The effective masses
reported in the AFLOWLIB database are averages computed
taking into account
all symmetry considerations and contributions from bands falling within 26 meV of the band
edges \cite{aflowlib_2,aflowlib_3}, providing thus a
reliable estimate. With respect to the band gap, we note that
the value reported in the AFLOWLIB database is semiempirical,
resulting from a
combination first-principles computations and a
least-squares fit to the experimental gaps of a set of 100 selected compounds.
Compared to experiment,
the reported fitted values present
a percentage error root-mean-square of 24\% \cite{aflowlib_3},
which is quite reasonable for such a large high-throughput database. However,
because of the margin of error, and to try
to avoid discarding oxides that experimentally
might in fact be transparent in the whole visible range,
we set a lower limit of 2.5 eV for the fitted band gap
to accept an oxide in our selection at this stage.

The above criteria
result in a list of 2 binary oxides, 41 ternaries, and 27 quaternaries for which to proceed with higher accuracy
electronic structure calculations in the next stage. However,
we found that among these there are numerous
oxides presenting magnetic order
(ferromagnetic or antiferromagnetic).
These entail calculations that are more involved and
heavier than in the case of non-magnetic oxides.
To reduce the number of such calculations,
we screen out those oxides with magnetic order that at the
same time contain elements that
might raise toxicity concerns (i.e., As, Cd, Hg, Tl, Pb, and F).
FeTeFO$_3$ is screened out because its synthesis procedure likely involves toxic fluorine gas.
Indeed, these are probably
of less interest from an industrial perspective.
Consequently, the list of oxides considered for
further calculations consists of 2 binaries, 37 ternaries, and 21 quaternaries, i.e., 60 oxides in all.
These oxides
are catalogued in Fig.~\ref{fig1}, with information discussed in the
next Subsection. (See Supplementary Table S1 for the complete list of oxides,
i.e., including those with potentially toxic elements, where we
report the structural parameters for each compound in addition
to their band gap values and effective masses.)

\begin{figure*}
\centering
\includegraphics[width=0.75\hsize]{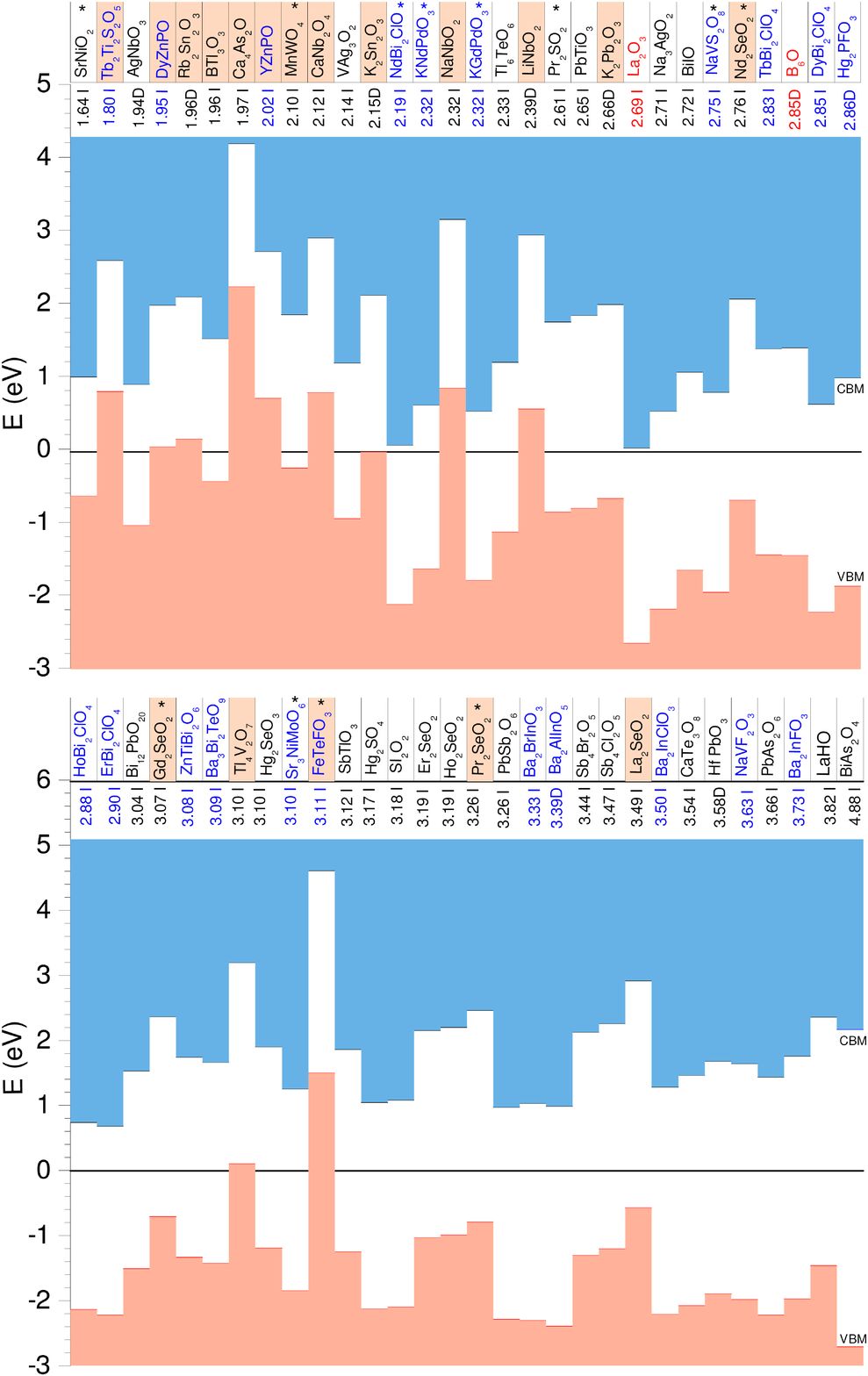}
\caption{\label{fig1} {\bf Valence and conduction band
energies with respect to the
branch point energy.}
The oxides are arranged in ascending order according to their HSE06 band gap value, a ``D'' (``I'') indicating a direct
(indirect) fundamental gap. The highlighted oxides are those
that are easily doped $p$-type, according to our criterion
(see text). Quaternary oxides are written in blue, ternaries
in black, and binaries in red.
}
\end{figure*}

\subsection*{$p$-type dopability screening}

Determining the $p$-type
(or $n$-type) dopability of a semiconductor
using first-principles methods requires that the physics
of defects is reliably accounted for and, accordingly, the electronic
structure of the semiconductor, including the band gap
value. Indeed, the $p$-type dopability of a semiconductor depends
on whether acceptor-like defects are easily formed and
whether it is prone to the spontaneous generation of compensating
native defects (e.g., anion vacancies). Thus, the well known
problem of the band gap underestimation by exchange-correlation
functionals typically used in first-principles studies,
such as the local density approximation (LDA) or the generalized
gradient approximation (GGA), has been a hurdle until
recently. Thanks to notable methodological developments
in recent years, however, first-principles methods are capable
today of a broader and deeper description of the physics of defects
\cite{freysoldt14}, albeit at a greater computational cost. In the case of
studies based on the Heyd, Scuseria, Ernzerhof (HSE06) hybrid-functional \cite{HSE1,alpha_HSE}, for instance,
computational cost increases typically by an order of magnitude,
or more, compared to the approximations just mentioned.
Nevertheless, because of its reliability, the HSE06
hybrid-functional has practically become the functional of
choice in the study of defects \cite{freysoldt14},
and is the one we use
here (see the Methods section for further details).

A criterion with significant predictive power,
indicating whether a wide band gap material
is easily $n$- or $p$-dopable,
is provided by the branch point energy (BPE) or charge neutrality level \cite{king08,robertson11,scanlon13}.
This is the energy level below which defect
states in the gap will have a mainly donor-like, and above
which they will have a mainly acceptor-like character. This
is to say that it is the energy below (above) which the formation
of donor (acceptor) defects becomes favorable. Hence, in
a material with a BPE lying in the conduction band, or high in
the band gap, donor impurity defects will tend to be shallow,
while acceptor impurities will tend to be deep. Furthermore,
it is also more likely that there will be native donor defects
with very low formation energies for Fermi levels close to the
VBM. As a consequence, even if a shallow acceptor impurity
(or impurity complex) can be found, it will be difficult
for it to have a sufficiently low formation energy as to avoid
compensation by such native donors, which will act as killer
defects (this is exactly the problem with ZnO, for which the
BPE lies in the conduction band). Thus, to favor
$p$-type doping, the BPE should lie instead below the midgap,
and the lower the BPE, the easier $p$-type doping will be.
To determine the BPE value
we follow here the approach of
Schleife and co-workers \cite{21,hinuma14}, in which the BPE
is calculated as the
weighted average of the midgap energies over the Brillouin zone
(cf.~Methods for more details). The BPE depends solely on the
bulk band structure, which makes it a
computationally efficient screening tool.

In Fig.~\ref{fig1} we present the valence and conduction
band energies of
the oxides selected in the previous Subsection
with respect to the BPE, for ease of comparison.
The HSE06 calculated band gap values are
given
(a ``D''  or ``I'' indicating a direct or indirect fundamental
gap), and the oxides are sorted in ascending order according
to their gap value. Quaternaries
are indicated in blue, while binaries are indicated in red.
The position of the BPE with respect to the band edges allows
us to classify the studied oxides in three classes. In the first
class we have the oxides that in this work we consider to be
easily $p$-type dopable,
namely those for which the BPE lies in the valence band or
in the lower fourth of the gap above the VBM. Oxides with a
BPE above this limit might still be $p$-type dopable, but
with our choice we try to ensure an easy $p$-type dopability.
The compounds in this first class are highlighted in Fig.~\ref{fig1}.
There are fourteen ternary and four quaternary oxides in this class.

In the second class
we classify the oxides for which the BPE lies above the CBM, or in the upper fourth of the gap below the CBM. These oxides are easily doped $n$-type.
We do not focus on $n$-type TCOs in this work, and we just mention that La$_2$O$_3$, Na$_3$AgO$_2$ and the quaternaries KGdPdO$_3$, and Bi$_2$ClXO$_4$, where X$=$Dy, Ho, Nd, or Er, belong to this class. Note that
well known $n$-type TCOs, such as
In$_2$O$_3$ or ZnO, did not make it to our list of oxides
in Fig.~\ref{fig1}
because they present a hole mass larger than 1 $m_e$.

The third class consists of the rest of the oxides. Oxides for
which the BPE is closer to the center of the band gap than to
the band edges do not follow any obvious trend. Some may
still be $n$- or $p$-type dopable, and some may be both (ambipolar). However, for oxides with a band gap, such as those considered here, this is probably difficult. Indeed, ambipolar
doping becomes increasingly challenging the wider the band
gap of a material because
the electron affinity is becoming small and/or the ionisation
energy is becoming
large \cite{zunger03,robertson11,yan08}.
(Conversely, ambipolar doping is easy in narrower gap
semiconductors. Si and Ge are among the best known examples.
\cite{ashcroft76}.)

\begin{table}[h]
\centering
  \caption{\label{table2}
{\bf Properties of the easily $p$-type dopable,
low hole effective mass, transparent oxides identified via
the selection procedure in this work.}
The fundamental band gap, $E_{g}$, first direct band gap,
$E_{g}^d$, second gap direct gap in the valence band,
$E_{g,VB}^2$, and standard enthalpy of formation,
$\Delta H_f$,
are our HSE06 calculated values. The average hole
effective mass ($m^*_h$) is the AFLOWLIB value.
(Energies are in eV, and effective masses are in units of $m_e$.)}
  \begin{tabular*}{0.48\textwidth}{@{\extracolsep{\fill}}lccccl}
      \hline
    oxide& $E_g$ & $E_g^d$ & $E_{g,VB}^2$\footnote{For the compounds that have a non-zero total magnetic moment $E_{g,VB}^2$ is the energy difference between the two highest occupied bands with the same spin component.} & $m^*_h$ & $\Delta H_f$\footnote{Enthalpy of formation energy per formula unit
with respect to the constituent elements in their standard
phases.}\\\hline
    La$_2$SeO$_2$ & 3.49 & 4.02 & 1.55 & 0.92 & -15.62 \\
    Pr$_2$SeO$_2$ & 3.26 & 4.09 & 1.99 & 0.69 & -15.09 \\
    Nd$_2$SeO$_2$ & 2.76 & 3.12 & 1.76 & 0.79 & -14.72 \\
    Gd$_2$SeO$_2$ & 3.07 & 3.95 & 2.28 & 0.76 & -32.67 \\
     \hline
     \end{tabular*}
\end{table}

\subsection*{$p$-type dopable, low hole mass, transparent oxides}

As mentioned above, the AFLOWLIB band gap values
present a margin of error, and it is important to verify
the magnitude of the band gaps using a more accurate
method.
We find that while none of the
$p$-type dopable oxides has an underestimated band gap in the
AFLOWLIB database (this is the case for a few of the oxides in
the other classes, such as BiAsO$_4$ or ZnTiBi$_2$O$_6$;
cf.~Supplementary Table S1), there are several for which
the gap is overestimated. Thus, for instance, LiNbO$_2$ and
K$_2$Pb$_2$O$_3$ have direct band gaps that in the AFLOWLIB
database are reported to be of
3.16 eV and 3.20 eV, respectively, but the corresponding HSE06
values are of 2.39 eV and 2.66 eV, respectively.
Hence, although these oxides are $p$-type
dopable and have a low hole effective mass, their complete
transparency is not ensured. (Note that our finding
regarding K$_2$Pb$_2$O$_3$ agrees quite well with
reference \citenum{7b}.)
On the other hand, one must take care of not discarding
too rapidly an oxide with a low indirect fundamental gap,
because it
is direct transitions that are most
important for transparency. Therefore,
we screen our $p$-type dopable oxides searching for those
with a HSE06 direct band gap band gap $> 3.1$ eV. This is
sufficient to ensure their nominal
transparency in all the visible range.
Indeed, if the direct band gap is symmetry-forbidden, the
direct-allowed band gap will {\it a fortiori} be larger.
Discarding oxides containing elements that might pose
toxicity problems,
this results in the list of four
oxides in Table~\ref{table2},
X$_2$SeO$_2$, with X = La, Pr, Nd, and Gd.
These oxides present excellent characteristics for
$p$-type TCO applications, and are completely novel as such.
For completeness, Table~\ref{table2} includes
the HSE06 values for the fundamental gap, the direct gap, the so-called second gap (we come back
to this point in the Discussion section), the
average hole mass reported in the AFLOWLIB, and the
enthalpy of formation.
We note that the last three oxides possess ferromagnetic order
(for details, see Ref.~\citenum{aflowlib}), which offers the opportunity to explore
further applications of $p$-type TCO materials.

\subsection*{$p$-type La$_2$SeO$_2$}

\begin{figure}
\centering
\includegraphics[width=0.8\hsize]{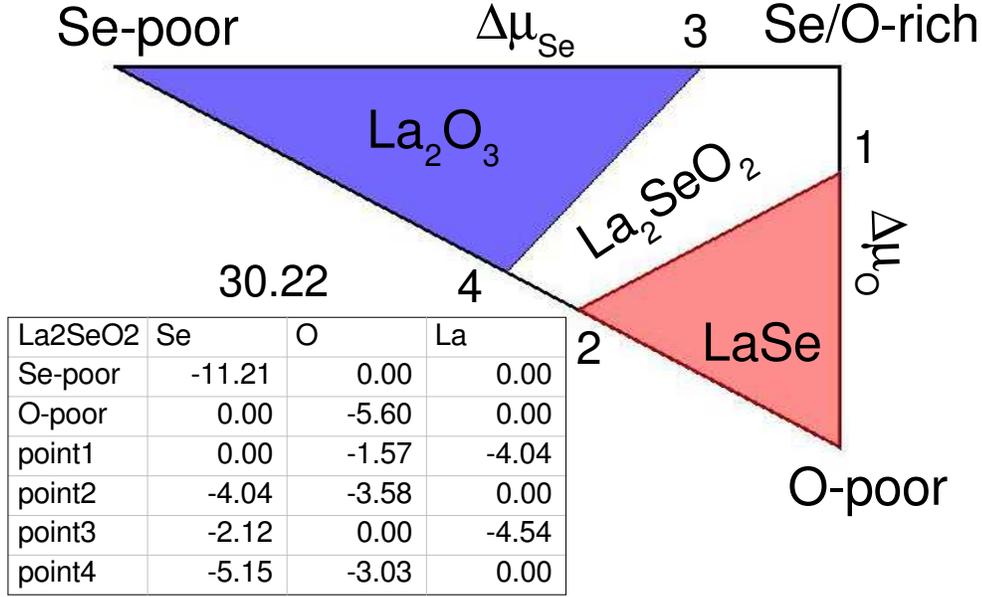}
\caption{\label{fig2} {\bf La$_2$SeO$_2$ stability triangle
in the chemical potentials plane.}
The white area defines the range of chemical potentials
in which La$_2$SeO$_2$
 is stable against precipitation of competing
binary phases La$_2$O$_3$ and LaSe.
The chemical potentials are given with respect
to their standard phases, i.e., $\Delta\mu_i=\mu_i-\mu_i^0$, where $\mu_i^0$ corresponds to the solid metal for
selenium and to the diatomic molecule for oxygen.}
\end{figure}

To corroborate the assertion that the oxides in our final list
are $p$-type dopable, we consider the case of La$_2$SeO$_2$
and show explicitly that Na will act as a shallow acceptor
with charge state $2-$ and that anion vacancies will not act
as hole killers in suitable growth conditions.
The formation energy of a defect $D$, in charge state $q$,
in a bulk compound is given by \cite{vandewalle04}
\begin{equation}\label{formation}
E_f[D^{q}]=E_{\rm tot}[D^q]-E_{\rm tot}[{\rm bulk}]
-\sum_i n_i\mu_i+q[E_F+E_v+\Delta V].
\end{equation}
In the above,
$E_{\rm tot}[D^q]$ is the total energy of defect-containing
system and
$E_{\rm tot}[{\rm bulk}]$ is the total energy of the defect-free system,
$n_i$ is the number of atoms of type $i$ added to or removed from
the system ($n_i<0$ if the atom is removed and $n_i>0$ if the atom is added), with
$\mu_{i}$ the corresponding chemical potential. $E_F$
is the electronic
chemical potential, measured with
respect to the VBM, $E_v$, of the undoped system.
$\Delta V$ is a reference alignment term (see
the Methods section for further details).

\begin{figure*}
\centering
\includegraphics[width=0.75\hsize]{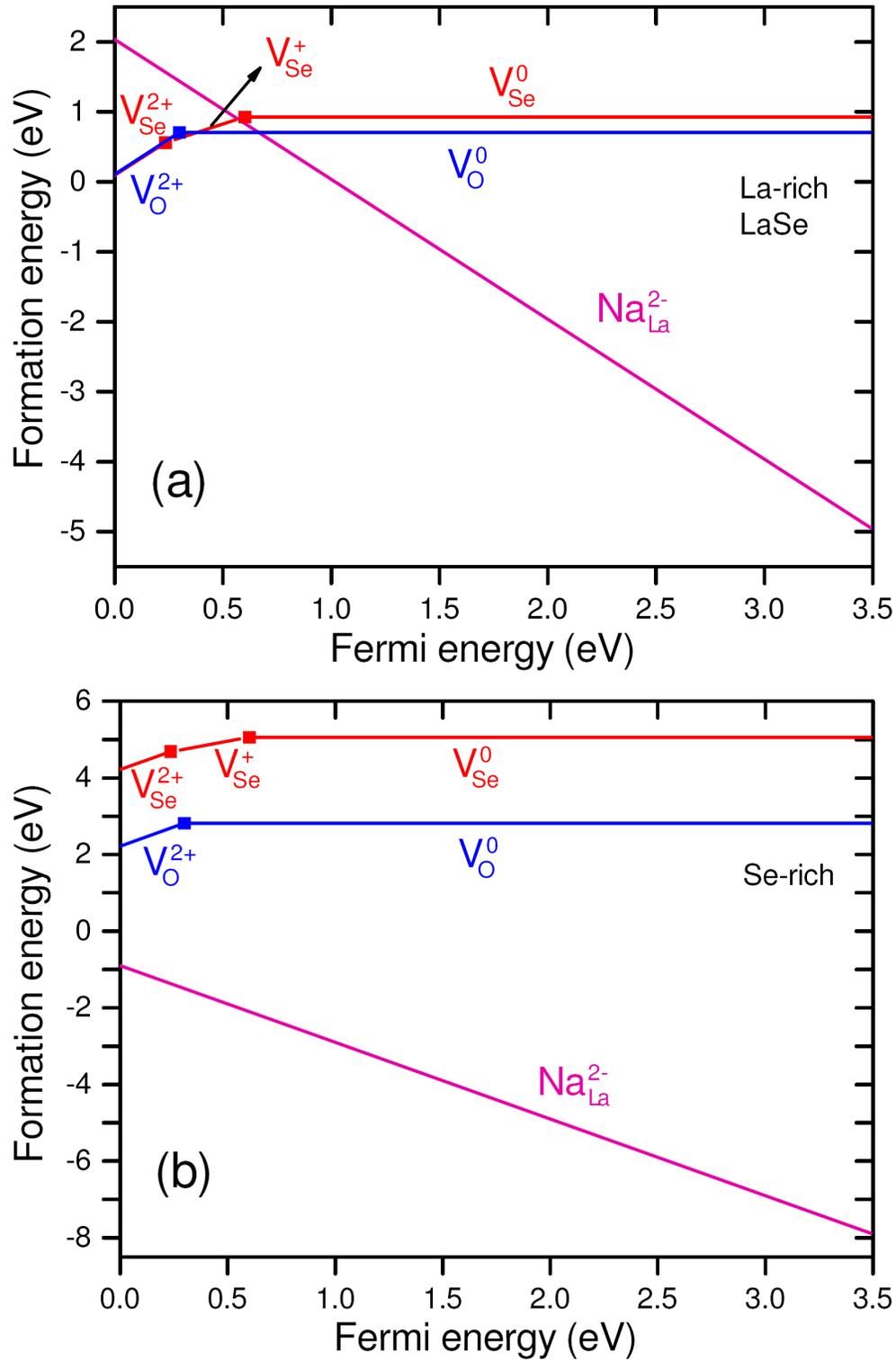}
\caption{\label{fig3} {\bf Formation energy of Na$_{\rm La}$,
$V_{\rm Se}$ and $V_{\rm O}$ as a function of Fermi energy.}
For all defects only
the charge state with lowest formation energy is shown at each
Fermi energy value. ({\bf a}) In La-rich
conditions, Na$_{\rm La}^{2-}$ is a shallow acceptor, but will
tend to be compensated by anion vacancies ($V_{\rm Se}$ and
$V_{\rm O}$). ({\bf b}) In
Se-rich conditions Na$_{\rm La}^{2-}$ will not be compensated
by anions vacancies and effectively dope
La$_2$SeO$_2$ $p$-type.
}
\end{figure*}

The formation energy of a defect
depends quite importantly on the chemical
potentials of the atomic species involved. The relevant
range of $\mu_i$ values is delimited by the stability of
La$_2$SeO$_2$ against the precipitation of competing
binary compounds La$_2$O$_3$ and LaSe. The corresponding
stability triangle is shown in Fig.~\ref{fig2}.
Since La
is expected to be in nominal oxidation state $+3$ in
La$_2$SeO$_2$, we consider Na$_{\rm La}$ as a possible
acceptor impurity. (In principle, one could consider K
impurities as well. However, the atomic radius of K is
larger than the one of La \cite{slater64},
and so K$_{\rm La}$
will tend to have higher formation energies.)
We also consider oxygen and
selenium vacancies ($V_{\rm O}$ and
$V_{\rm Se}$) as
possible compensating donor defects. In Fig.~\ref{fig3}
we show the formation energies of the three defects,
as a function of $E_F$, in
La-rich [Fig.~\ref{fig3}(a)], with chemical potentials
corresponding to point 2 in Fig.~\ref{fig2}, and Se-rich conditions
[Fig.~\ref{fig3}(b)], with chemical potentials corresponding
to point 1 in Fig.~\ref{fig2}. Only the formation energy of
the lowest charge state is shown at any given $E_F$.
Fig.~\ref{fig3}(a) shows that Na$_{\rm La}$ is a shallow
acceptor in charge state $2-$, but that it will be compensated
by $V_{\rm Se}$, which acts as a deep donor. On the other
hand, Fig.~\ref{fig3}(b) shows that in Se-rich conditions
(and moderately O-rich conditions), Na$_{\rm La}^{2-}$ will
tend to form spontaneously and that it will
not be compensated by either $V_{\rm O}$ or $V_{\rm Se}$.
Note that taking the chemical potentials corresponding to
point 3 in
Fig.~\ref{fig2}, will just revert the order of the
formation energy curves for $V_{\rm O}$ and $V_{\rm Se}$.
This shows that
Na-doped La$_2$SeO$_2$ will indeed behave as a stable $p$-type
TCO in strong to moderate anion-rich conditions.

\section*{Discussion}

The merit of our list of oxides in Table~\ref{table2},
which present
direct band gaps ranging from 3.12 eV to 4.09 eV, and
hole effective masses ranging from 0.69 $m_e$ to 0.92 $m_e$,
is readily recognised by comparing these with the
properties of current $p$-type TCOs. Among the $p$-type
oxides reported by Wager and co-authors
in Ref.~\citenum{wager08}, Sr-doped LaCuOS, with a
direct band gap estimated to be 3.1 eV, presents one of
the highest optical transmissions \cite{ueda00}.
Note that its band gap value is in the low end of the
values in our list of oxides.
The $p$-type conductivity of LaCuOS,
on the other hand, suffers from a low mobility.
Mobility is improved by partially replacing S with Se,
raising it from below 1 cm$^2$V$^{-1}$s$^{-1}$ up to
$\sim 8$ cm$^2$V$^{-1}$s$^{-1}$ \cite{hiramatsu03}, one of the
best mobilities in the list of Wager and co-authors
\cite{wager08}.
But transparency degrades in the process, as the band gap
decreases with Se content down to a vale of 2.8 eV for
LaCuOSe \cite{hiramatsu03}. Interestingly, the hole effective
mass in LaCuOSe was experimentally estimated to be
of $\sim 1.6$ $m_e$ \cite{hiramatsu07}, which is appreciably
higher than the corresponding values in our
Table~\ref{table2}. We conclude
from the above that one can indeed expect that a suitable synthesis
or growth procedure applied to the oxides in Table~\ref{table2}
will result in better $p$-type TCOs, regarding both
transparency and hole conductivity.

The $p$-dopability of the oxides in Table~\ref{table2} can be
readily understood in the light of previous work. Indeed, the oxyselenides can be viewed as further examples of the idea
of Hosono and co-workers of exploiting the stronger delocalisation of the $p$-orbitals of chalcogens to raise the VBM energy
and facilitate $p$-doping \cite{hosono07}. Of course, this is
also the reason for their low hole mass.

In Table~\ref{table2} we indicate the value of the
second band gap, which in a $p$-type TCO is defined
as the energy difference between VBM and the next
eigenvalue below it. This might be of interest in the case of
heavily doped TCOs, because at sufficiently high carrier
concentration the absorption of photons by the latter will
start to be favored and will tend to limit the
transparency \cite{mryasov01}. However, in active electronic
TCO applications (as opposed to passive electronic TCO
applications), rather low carrier concentrations are sought
\cite{wager08}, and in that case a low second gap is not
an issue.

We mentioned above that in Ref.~\citenum{7b} a high-throughput
search of low hole mass, wide band gap oxides is performed to
identify $p$-type TCO candidates. There is no overlap between the good candidates in Ref.~\citenum{7b} and our final list basically because of the different population of oxides studied and because the search criteria are different.
Indeed, in Ref.~\citenum{7b} no rare-earth containing oxides
were considered, and no quaternaries. This excludes the oxides in our Table~\ref{table2} from their list.
On the other hand, of the candidates presented in Ref.~\citenum{7b}, ZrOS and
Na$_2$Sn$_2$O$_3$ are not in our list in Table~\ref{table2} because of their hole effective masses are higher than 1 $m_e$. PbTiO$_3$ and B$_6$O do not make it into our list because they do not fall in our class of easily $p$-type dopable oxides. This is in line with a recent first-principles report, which indicates that oxygen vacancies in PbTiO$_3$ will tend
to act as hole-compensating defects \cite{shimada13}.
With respect to B$_6$O, which falls in the third class of oxides according to our classification, a recent study on its possible ambipolarity has shown that it indeed will not be easy to dope $p$-type \cite{B6O}. The only possible $p$-type dopant identified for B$_6$O with low enough formation energy near the VBM is a (CH)$_{\rm O}$ complex but it requires at the same time that
substitutional carbon in boron position (C$_{\rm B}$) defects are avoided. These results confirm our conclusions regarding their dopability.
Finally, K$_2$Sn$_2$O$_3$ and K$_2$Pb$_2$O$_3$ are
excluded because their HSE06 direct band gaps are lower than 3.1 eV.

To summarise, in this work we perform a search of new $p$-type TCOs
following a
high-throughput based on first-principles methods.
We first screen all the binary, ternary and quaternary oxides in the AFLOWLIB database to identify those compounds that are reported to have a band gap larger than 2.5 eV, and a hole mass lower than 1 $m_e$. We calculate the electronic structure of the thus identified compounds using state-of-the-art
methods
in order to determine their $p$-type dopability via the position of their BPE with respect to the band edges. We
further require the oxides to have a direct gap larger than 3.1 eV aiming at
ensuring their transparency in all the
visible energy range. The list of $p$-type dopable, low hole
effective mass, transparent oxides that we obtain consists of
La$_2$SeO$_2$, Pr$_2$SeO$_2$, Nd$_2$SeO$_2$, and Gd$_2$SeO$_2$. Furthermore,
we explicitly show that in suitable growth
conditions Na impurities will behave as
shallow acceptors in La$_2$SeO$_2$, when substituting La, and
that anion vacancies will not compensate them.
Because of their
characteristics, these oxides have the potential to outperform the currently used $p$-type TCO materials and to lead to
a breakthrough in transparent electronics applications.
Indeed, to our knowledge none of the stable
$p$-type TCOs reported so far in the literature
has simultaneously a hole
effective mass $< 1$ $m_e$ and a band gap $>3.1$ eV.
We hope that experimentalists will be intrigued by
our results and will be encouraged to
try to confirm our findings.

\section*{Methods}

\subsection*{First-principles calculations}

All structural and electronic properties calculations in this work
are performed within density functional theory (DFT)
\cite{Hohenberg_Kohn1,kohn_sham_1965},
using the plane-wave basis sets and the projector
augmented-wave method \cite{blochl94}, as implemented in the
Vienna Ab-initio Simulation Package
(VASP)\cite{c3,10b,vasp,ch2_78}.
We use the Heyd, Scuseria, Ernzerhof (HSE06) hybrid functional for the exchange-correlation potential \cite{HSE1,alpha_HSE}
(with the standard 25\% of exact echange) to calculate
the lattice parameters and to relax the atomic positions, as well as to determine the electronic structure and to
determine formation energies.

As discussed by Jain {\it et al.} \cite{jain11}, in a high-throughput approach, it is impractical to perform rigourous energy cutoff and {\bf k}-point convergence studies during the screening procedure. Thus, we use an energy cutoff of 400 eV for the plane-wave basis set, which is sufficiently high to ensure convergence. To sample the Brillouin zone, we use a Monkhorst-Pack grid \cite{MP}, making sure that the $\Gamma$ point is included in the mesh. For the number of {\bf k}-points in the mesh, we follow Ref.~\citenum{jain11}, sampling the first Brillouin Zone using a grid of at least $500/n$ {\bf k}-points, where $n$ is the number of atoms in the unit cell. The convergence tests are performed for the candidates selected at the end of the screening procedure, focusing on the BPE and formation energy. Finally, we note that atomic relaxations are made until residual forces on the atoms are less than 0.01 eV/{\AA} and total energies are converged to within 1 meV.

\subsection*{Branch point energy}

As indicated above,
the BPE is calculated as a weighted average of the midgap energies over the Brillouin zone \cite{21,hinuma14},
\begin{equation}\label{eq1}
E_{BP}=\frac{1}{2N_{k}}\sum_k[\frac{1}{N_{CB}}\sum_i^{N_{CB}}\varepsilon_{c_{i}}(k)+\frac{1}{N_{VB}}\sum_i^{N_{VB}}\varepsilon_{v_{j}}(k)].
\end{equation}
Here, $N_k$ is the number of points in the {\bf k}-point mesh, $N_{CB}$ and $N_{VB}$ are the number of conduction and valence bands considered for the averaging, with $\varepsilon_c$ and $\varepsilon_v$ their corresponding energies. The {\bf k}-point grid is sufficient to have a $E_{BP}$ converged with respect to the number of {\bf k}-points. The number of valence and conduction bands used is determined by scaling them according to the number of valence electrons in the primitive cell (excluding d electrons), as in the work of Schleife {\it et al} \cite{21}. Note that in the latter work it is recognised that the dependence of $E_{BP}$ on the number of bands used results in an uncertainty of $\sim 0.2$ eV on its value. However, given that we consider here wide band oxides
($E_g\leq 2.5$ eV), and
that we impose a strong criterion, namely that the $E_{BP}$ should not lie above the VBM by more than 1/4 of $E_g$,
the uncertainty mentioned will not change our conclusions regarding the easy dopability of our selected compounds.

\subsection*{Defect formation energy and stability triangle}

To model the
defect system we use a 90-atom $3\times3\times2$ supercell, and sample the
Brillouin zone using a $2\times2\times2$, $\Gamma$-centered,
Monkhorst-Pack grid.
We denote
the chemical potential of the constituent elements $\mu_{\rm La}$, $\mu_{\rm Se}$, and $\mu_{\rm O}$.
The chemical potentials are related to the enthalpy of formation of the oxide through $\Delta H_f({\rm La}_2{\rm SeO}_2)=2\Delta\mu_{\rm La}+\Delta\mu_{\rm Se}+2\Delta\mu_{\rm O}$. As
indicated previously, $\Delta\mu_i=\mu_i-\mu^0_i,$ with $\mu^0_i$ the chemical potential in the standard phase of the
element, i.e., metallic La and Se, and molecular O (note
that a spin-polarised calculation must be performed in the
latter case). In a first instance, the range of possible chemical potential values are determined by the value of the enthalpy of formation and the limits imposed by the precipitation of the constituent elements, i.e., $\Delta\mu_{\rm La}=0$ (La-rich), $\Delta\mu_{\rm Se}=0$ (Se-rich), and $\Delta\mu_{\rm O}=0$ (O-rich). This results in the triangular area in the $(\Delta\mu_{\rm Se},\Delta\mu_{\rm O})$ plane
plotted in Fig.~\ref{fig2}, which is known as a stability triangle in the literature \cite{29a}. As indicated in
Fig.~\ref{fig2}, the relevant range of chemical potentials
is further limited by phase segregation, i.e., the enthalpy of formation of binary oxides, $\Delta H_f({\rm La}_2{\rm O}_3)=2\Delta\mu_{\rm La}+3\Delta\mu_{\rm O}$ and $\Delta H_f({\rm LaSe})=\Delta\mu_{\rm La}+\Delta\mu_{\rm Se}$. This defines the area within the stability triangle, in which ${\rm La}_2{\rm Se}{\rm O}_2$ is stable, and the exact chemical potential
values of, e.g., points 1 and 2. These are feeded to
Eq.~(\ref{formation}) to calculate
the formation energies plotted in Fig.~\ref{fig3}in the main article. Note that in Eq.~(\ref{formation}) the VBM of the undoped oxide is used as
reference ($E_v$) for the electronic chemical potential, and
the alignment term ($\Delta V$) is calculated following the
procedure introduced in
Ref.~\citenum{saniz13} (see also
Ref.~\citenum{dabaghmanesh13}).

\bibliography{ptype}

\section*{Acknowledgments}
We acknowledge the financial support of FWO-Vlaanderen
through project G.0150.13 and of a GOA fund from the University
of Antwerp. The computational resources and services
used in this work were provided by the VSC (Flemish Supercomputer
Center) and the HPC infrastructure of the University
of Antwerp (CalcUA), both funded by the Hercules Foundation
and the Flemish Government–department EWI.

\section*{Author contributions}
N.S. performed the simulations; R.S. and N.S. wrote the manuscript; R.S., B.P., and D.L. coordinated the work; All authors reviewed and commented on the manuscript.

\section*{Additional information}
{\bf Supplementary Information} is available at http://www.nature.com.

\noindent
{\bf Competing financial interests}: The authors declare no competing financial interests.

\noindent
{\bf Reprints and permission} information is available online at
http://npg.nature.com/reprintsandpermissions/

\noindent
{\bf Journal Reference} Scientific Reports 6 (2016) 20446

\noindent
{\bf DOI} 10.1038/srep20446
\end{document}